\documentclass[10.5pt]{article}

\usepackage{color,graphicx}
\usepackage{young}
\usepackage[vcentermath]{youngtab}
\usepackage{amsmath,amssymb,graphicx,multirow}
\usepackage[bookmarks=true,colorlinks=true,linkcolor=blue,citecolor=blue,urlcolor=blue,bookmarksnumbered]{hyperref}
\definecolor{darkred}{rgb}{0.65,0.15,0}
\hypersetup{pdfborder={0 0 0},colorlinks=true,urlcolor=darkred,citecolor=blue,linkcolor=darkred,linktocpage=true}

\newcommand{\arXivlink}[1]{\href{http://arXiv.org/abs/#1}{arXiv:#1}}

\usepackage{cite}
\usepackage{amsmath}
\usepackage{amsfonts}
\usepackage{amssymb}
\usepackage{graphicx}%
\usepackage{amsthm}
\usepackage{mathrsfs}
\usepackage[T1]{fontenc}
\usepackage{enumerate}
\usepackage{graphicx}
\usepackage{multirow}
\usepackage{amsmath,amssymb,amsfonts}
\usepackage{amsthm}
\usepackage{mathrsfs}
\usepackage[title]{appendix}
\usepackage{xcolor}
\usepackage{textcomp}
\usepackage{manyfoot}
\usepackage{booktabs}
\usepackage{algorithm}
\usepackage{algorithmicx}
\usepackage{algpseudocode}
\usepackage{listings}

\usepackage{color}
\theoremstyle{thmstyleone}%
\newtheorem{theorem}{Theorem}
\theoremstyle{thmstyletwo}%

\theoremstyle{thmstylethree}%
\def\4diml{four-dimensional}

\def\-1{^{-1}}

\newcommand{\M}{\mathscr{M}}

\newcommand{\G}{\mathscr{G}}

\makeatletter

\@addtoreset{equation}{section}
\makeatother
\begin{document}

\thispagestyle{empty}

\vspace{5mm}

\begin{center}
{\Large \bf Yang-Baxter deformations of the OSP$(1|2)$ WZW model}

\vspace{10mm}

\textrm{Ali Eghbali\footnote{\texttt{Author to whom any correspondence should be addressed.\\
eghbali978@gmail.com\\ yaghobsamadi@gmail.com\\ rezaei-a@azaruniv.ac.ir }}, Yaghoub Samadi and Adel Rezaei-Aghdam
\\}

\vspace{0em}

{\it Department of Physics, Faculty of Basic Sciences,\\
Azarbaijan Shahid Madani University, 53714-161, Tabriz, Iran}
\end{center}

\vspace{0em}

\begin{abstract}\noindent 

\noindent
We obtain inequivalent classical r-matrices of the $osp(1|2)$ Lie superalgebra as real solutions
of the graded (modified) classical Yang-Baxter equation,
in such a way that the corresponding automorphism transformation is employed.
Then, Yang-Baxter deformations of the Wess-Zumino-Witten model based on the OSP$(1|2)$ Lie supergroup
are specified by super skew-symmetric classical r-matrices.
In this regard, the effect coming from the deformation is reflected as the coefficient of both metric and $B$-field.
Furthermore, it is shown that all resulting classical r-matrices are non-Abelian and also non-unimodular,
which leads us to graded generalized supergravity equations.
We show that the background of undeformed model is a solution of the graded generalized supergravity equations when supplemented
by an appropriate supervector field obtaining from the linear combination of the Killing supervectors corresponding to the background,
while the deformed models do not satisfy these equations. This is consistent with our expectations,
since the deformed models under consideration do not describe a
Green-Schwarz superstring.
However, the deformed backgrounds are interesting, in particular in the OSP$(1|2)$ case they are rare, much hard technical work was needed to obtain them.
\end{abstract}

\vspace{1mm}

\setcounter{page}{1}

\section{\label{Sec.I} Introduction}

In the past decade, significant interest has been drawn towards Yang-Baxter (YB) deformations of $\sigma$-models because
the recent motivation for studying these models has come from their applications to string theory and AdS/CFT.
These models which were originally defined for
the principal chiral model on a simple group \cite{{Klimcik1},{Klimcik2}}, were built on classical r-matrices (CrMs)
that solve the classical Yang-Baxter equation (CYBE).
There is a variety of CrMs, providing various deformations of integrable $\sigma$-models.
CrMs as the initial input for construction of the YB $\sigma$-models may be divided into Abelian and non-Abelian.
It has been shown that the YB deformed chiral models related to Abelian and non-Abelian CrMs correspond to T-duality
shift T-duality transformations \cite{Osten} and deformed T-dual models \cite{Borsato1}, respectively.
However, studying integrable deformations is one approach to achieving solvable string theories, which is always interesting.

An important application of the YB $\sigma$-model description is the integrable deformation of the $AdS_5 \times S^5$ superstring
based on the solutions of the modified (m)CYBE \cite{Delduc2} and CYBE \cite{Kawaguchi1},
resulting in a quantum deformation of the superconformal symmetry of $AdS_5 \times S^5$ \cite{Delduc3}.
The metric and $B$-field corresponding to the YB deformed background obtained from the solutions of the mCYBE were driven in \cite{Arutyunov1},
and the full background was then studied in \cite{{Arutyunov2},{Hoare1}}
(see, also, \cite{{Tongeren},{HidekiKyono},{Hoare.van Tongeren},{Yoshida2},{Tongeren2}}).
Later, the YB $\sigma$-model was generalized by adding a WZW term \cite{{Delduc4},{Klimcik2017},{Quantum},{2020},{N. Mohammedi},{taghavi}}
(for a nice review, see \cite{Hoare}).
Various aspects of the YB deformed WZW models along with some examples were discussed in
\cite{{Yoshida.NPB},{Klimcik.LMP},{Epr1},{Epr3},{Daniele1},{Daniele2},{Daniele3}}.
In most of the works on the deformation of superstring action,
attention was focused on the case where the deformations were caused by the bosonic generators of the Lie supergroup,
until in \cite{Epr2}, the deformation was performed on both the bosonic and fermionic sectors of the models.
In \cite{Epr2}, the deformation prescription invented by Delduc {\it et al.} \cite{Delduc4} has been extended to the Lie supergroup case,
and some explicit examples based on the non-semi-simple Lie supergroups have been given.

In the present work, we investigate YB deformations of the WZW model based on the OSP$(1|2)$ Lie supergroup
by following a prescription generalized in \cite{Epr2}.
The simple Lie superalgebra $osp(1|2)$ is interesting as it plays in the supersymmetric case a role analogous to $sl(2, \mathbb{R})$.
We first calculate inequivalent CrMs for the $osp(1|2)$ as real solutions of the graded (m)CYBE.
It is worth stressing that in order to calculate Poisson brackets on the OSP$(1|2)$,
the authors of Ref. \cite{J.z} found three independent CrMs for the $osp(1|2)$. In this way, they
showed that all the corresponding Lie superbialgebra structures \cite{{N.A},{ER1}} were coboundaries.
Due to considering real parameters for CrMs, our results will differ from those of \cite{J.z}.
Note that in \cite{J.z}, the $osp(1|2)$ treated as complex.
However, as we will show, our CrMs lead us to five inequivalent families of the YB deformations of the OSP$(1|2)$ WZW model.
A quantum analogue of the $osp(1|2)$ and its finite-dimensional
irreducible representations were found in \cite{KULISH1}. There, to formulate the Hopf superalgebra of functions
on the quantum supergroup OSP$(1|2)$, the corresponding constant solution to the YBE has been used.
In addition, Kulish showed that \cite{KULISH2} a q-deformation of the graded
algebra $osp(1|2)$ can also be defined in relation with the graded YBE.
Later, by studying the quantum superalgebra of the $osp(1|2)$, the vertex models solution of the graded YBE and the associated interacting round a face models were constructed \cite{SALEUR90}.
Recently, the principal chiral model based on the OSP$(1|2)$ Lie supergroup has received attention \cite{Bielli1}.
There, the super non-Abelian T-duality of $\sigma$-models on group
manifolds with the emphasis on the T-dualization of $\sigma$-models whose target spaces are
supermanifolds \cite{ER2,ER5} has been performed \cite{Bielli1} (see, also, \cite{{Bielli2},{Bielli3}}).

According to \cite{Wulff1}, at the level of string theory, the
condition that the backgrounds of YB deformed models solve the standard supergravity equations of motion
requires the associated CrM to be unimodular.
As mentioned at the beginning of the Introduction, the CrMs may be crudely divided into two families:
Abelian and non-Abelian \cite{Osten}.
Abelian CrMs are always unimodular \cite{Wulff1}, meaning any such CrM maps a solution of standard supergravity to a solution of standard supergravity.
In contrast to Abelian ones, non-Abelian CrMs may be non-unimodular,
and indeed the associated backgrounds solve the generalized supergravity equations \cite{Arutyunov3}, but not the standard ones.
In \cite{Yoshida2}, by considering YB deformations of the $AdS_5 \times S^5$ superstring with non-Abelian CrMs
satisfying the homogeneous CYBE, it has been shown that the associated backgrounds satisfy the generalized supergravity equations.
In our case, we find that all CrMs of the $osp(1|2)$ are non-Abelian and also non-unimodular.
Unlike the YB deformations of the $AdS_5 \times S^5$,
the deformed backgrounds associated to the non-unimodular CrMs of the $osp(1|2)$ do not satisfy the graded generalized supergravity equations.
The fact that our deformed models do not satisfy these equations is consistent with expectations.
This is because the OSP$(1|2)$ $\sigma$-model is not a Green-Schwarz one.
The seminal work of Borsato and Wulff \cite{Wulff1}, which guarantees that $\eta $ and $\lambda$-deformations yield backgrounds satisfying the generalized supergravity equations, applies specifically to deformations of Green-Schwarz string models (see, also, \cite{Wulff3.revised}).
Our results are therefore in full agreement with the prior observation made by Bielli {\it et al.} \cite{Bielli1}
in the context of non-Abelian T-duality for the same OSP$(1|2)$ $\sigma$-model.
Bielli {\it et al.} showed that the non-Abelian T-dual backgrounds of the OSP$(1|2)$ $\sigma$-model did not satisfy the (generalized) supergravity equations.
They correctly identified the reason: the OSP$(1|2)$ $\sigma$-model is not a Green-Schwarz one, and therefore falls outside the scope of the Borsato-Wulff theorem.
The graded form of the generalized equations has recently been written in \cite{e.gh.r4}.
As an interesting example, we show that the undeformed background of the OSP$(1|2)$
WZW model satisfies the graded generalized equations when supplemented
by the supervectors $I$ and $Z$.

We have organized our manuscript as follows:
After the Introduction section, we start by introducing our notation,
and give a short overview of the YB deformation of WZW model based on a Lie supergroup $G$ in Section \ref{Sec.II}.
Section \ref{Sec.III} contains the original results of the work: in this section we
first introduce the simple Lie superalgebra $osp(1|2)$, and then build the WZW model based on the OSP$(1|2)$.
Furthermore, we solve the graded (m)CYBE to obtain the
$\mathbb{R}$-operators and inequivalent real CrMs for the $osp(1|2)$.
Finally, the YB deformed backgrounds of the OSP$(1|2)$ WZW model are calculated at the end of Section \ref{Sec.III}.
The discussion of solutions of the graded generalized supergravity equations with the undeformed background of the OSP$(1|2)$
WZW model and the YB deformed backgrounds of the model is devoted to Section \ref{Sec.IV}.
We conclude with a final discussion of the results with remarks and perspectives.
\section{Brief review of YB deformation of WZW model on a Lie supergroup}
\label{Sec.II}
As mentioned in the Introduction section, the generalization of YB deformation of WZW model
to a Lie supergroup $G$ case was performed in \cite{Epr2}.  Before proceeding to review the model, let us recall
the notations and some basic definitions that will be used throughout the paper.

\subsubsection*{{\it \underline{Notations and basic definitions}}}

A \emph{supervector space} ${V}$ is a ${\mathbb{Z}}_{2}$-graded vector space, a vector space as ${V}= { V}_{_B} \oplus {V}_{_F}$ over an arbitrary field $\mathbb{F}$ with a given decomposition of bosonic subspace ${V}_{_B}$ and fermionic one ${V}_{_F}$ with grades $0$  and $1$, respectively \cite{{Kac},{N.A}}.
The parity of a non-zero homogeneous element, denoted by $|x|$, is $0$ (bosonic) or $1$ (fermionic) according to
whether it is in ${ V}_{_B}$ or ${ V}_{_F}$, namely,
for any $x \in {V}_{_B}$ we have $|x|=0$, while $|x|=1$ for any $x \in {V}_{_F}$.
From now on, we work with the notation presented by Dewitt in \cite{D}.
In fact, we identify the indices grading by the same indices in the power of $(-1)$, i.e., we use $(-1)^x$ instead of  $(-1)^{|x|}$, where
$(-1)^x$ equals 1 or -1 if the subspace element is bosonic or fermionic, respectively.

A \emph{Lie superalgebra} ${\G}$ is a $\mathbb{Z}_2$-graded vector space, thus admitting the decomposition
${\G} ={\G}_{{_B}} \oplus {\G}_{_F}$, equipped with a bilinear
superbracket structure $[. , .]: {\G} \otimes {\G} \rightarrow {\G}$ satisfying the requirements of super anti-symmetry
and super Jacobi identity, respectively,
\begin{align}\label{2.1}
&[X , Y] =-(-1)^{XY}~ [Y , X],\nonumber\\
&[X , [Y , Z]] + (-1)^{X(Y+Z)}~ [Y , [Z , X]] + (-1)^{Z(X+Y)}~ [Z , [X , Y]]=0,
\end{align}
for any $X, Y, Z \in {\G}$ \cite{{Kac},{N.A}}. If ${\G}$ is finite-dimensional and the dimensions of ${\G}_{_B}$ and $ {\G}_{_F}$
are $m$ and  $n$, respectively, then ${\G}$ is said to have superdimension (m=d$_B$|n=d$_F$).
We also define $T_{_a}$'s, $a= 1, \cdots,$ sdim$\hspace{0.4mm}G$ as the bases in $\G$ of Lie supergroup $G$
which determine the (anti-)commutation relations
$[T_a , T_b] =f^c_{~ab}~ T_c$,
where $f^c_{~ab}$ are the structure constants of Lie superalgebra ${\G}$.
Additionally, we define a bilinear form $\big<.  , .\big>$ on ${\G}$ which is called supersymmetric iff
$\big<T_{a} , T_{b}\big>=(-1)^{ab}~\big<T_{b} , T_{a}\big>$,
and it is called super ad-invariant iff
$
\big<[T_{a} , T_{b}] , T_c\big> +(-1)^{ab}~\big<T_{b} , [T_{a} , T_c]\big> =0.
$

\subsubsection*{{\it \underline{WZW action based on a Lie supergroup $G$}}}

In order to define a WZW model on a Lie supergroup $G$, in general, given a Lie superalgebra ${\G}$ with generators $T_a,~$$a= 1, \cdots,$ sdim$\hspace{0.4mm}G$
and structure constants $f^c_{~ab}$, one needs a non-degenerate ad-invariant supersymmetric bilinear form $\Omega_{ab} = <T_a~ , ~T_b >$ on ${\G}$
so that it satisfies the following relation \cite{{nappi.witten},{ER7}}
\begin{eqnarray}\label{2.2}
f^d_{~ab} \;\Omega_{dc}+(-1)^{bc} f^d_{~ac} \;\Omega_{db} = 0.
\end{eqnarray}
The WZW model is a non-linear $\sigma$-model whose action is a functional of a field $g: \Sigma \rightarrow G$,
which is the embedding map of the two-dimensional worldsheet $\Sigma$ into the target supergroup $G$.
We introduce the basis $T_{_a}$ of $\G$, and the components of the
left-invariant Maurer-Cartan super one-forms $L_{\pm}=(-1)^a L_{\pm}^{a} T_{_a}=(-1)^a (g^{-1} \partial_{_\pm} g)^a T_{_a}$.
For future convenience we write the WZW action in terms of the left-invariant
super one-forms\footnote{Conventionally, this action comes with an additional factor of $\frac{K}{2 \pi}$ where $K$ stands for the the level of model,
and plays a crucial role in determining the symmetry algebra and the central charge of the model.
For simple supergroups the level has a well-defined meaning in the sense of multiplying a standard non-degenerate invariant form. There may also be quantization conditions linked to the topology of the supergroup.
We would like to include models whose metric renormalizes non-multiplicatively. Under these circumstances it is not particularly convenient to display the level explicitly and we assume instead that all possible parameters are contained in the metric.} \cite{ER7}
\begin{eqnarray}\label{2.3}			
S_{_{_{\hspace{0mm}WZW}}}(g)=\frac{1}{2}\int_{_\Sigma}d\sigma^+ d\sigma^-   (-1)^a L_{+}^{a} \Omega_{_{ab}} L_{-}^{b}  +\frac{1}{12}\
\int_{_{B_{3}}}d^{3}\sigma  (-1)^{a+bc}  \varepsilon^{\alpha\beta\gamma} L_{\alpha}^{a} \Omega_{ad}~ f^d_{~bc} ~ L_{\beta}^{b} L_{\gamma}^{c},~~
\end{eqnarray}
where $\sigma^\alpha = (\sigma^+ , \sigma^-)$ are the standard lightcone variables, and
$\varepsilon^{\alpha\beta\gamma}$ stands for the Levi-Civita symbol.
In the second term of \eqref{2.3} the integration is over the three-dimensional (super)manifold $B_3$ which has $\Sigma$ as a boundary.
Note that the WZW model \eqref{2.3} can be regarded as a $\sigma$-model in the form of
\begin{eqnarray}
S= \frac{1}{2}\int_{_{\Sigma}}\!d\sigma^+  d\sigma^- ~ (-1)^{\mu} \partial_{_+} x^{\mu} ({G}_{_{\mu \nu }}+{B}_{_{\mu \nu }}) \partial_{_-} x^{\nu},\label{2.4}
\end{eqnarray}
for which the line element and  Kalb-Ramond $B$-field may be, respectively, expressed as
\begin{eqnarray*}
ds^2 = (-1)^{\mu \nu} ~G_{_{\mu \nu}} dx^\mu~dx^\nu,~~~~~~~B = \frac{ (-1)^{\mu \nu}}{2} ~B_{_{\mu \nu}} dx^\mu  \wedge dx^\nu.
\end{eqnarray*}
Here $x^{\mu},$  $\mu= 1, \cdots,$ sdim$\hspace{0.4mm}G$ are local supercoordinates on the target supergroup
\footnote{The functions $x^{^\mu}$ include the
bosonic coordinates $x^i$ and the fermionic ones $\theta^{^\alpha}$,
and the labels $\mu, \nu$ run over $i =1,\cdots, d_{_B}$ and $\alpha = 1,\cdots, d_{_F}$, where $(d_{_B}|d_{_F})$ indicates
the superdimension of $G$.}.
We can therefore define the $H$-flux, a gauge-invariant closed three-form, $H=dB$,
such that\footnote{Here we have used the left partial differentiation to define $H=dB$.
According to \cite{D}, the relation between the left partial differentiation
and right one is given by
$$
{_{_{\mu}}{\overrightarrow \partial}} f := \frac{\overrightarrow{\partial}f}{{\partial} x^{^{\mu}}}  \;=\; (-1)^{\mu(|f|+1)}\;  \frac{f \overleftarrow{\partial}}{{\partial} x^{^{\mu}}},
$$
where $f$ is a differentiable function on ${\mathbf{R}}_c^m \times {\mathbf{R}}_a^n$ (${\mathbf{R}}_c^m$ are subset of all real numbers with dimension $m$ whereas ${\mathbf{R}}_a^n$ are subset of all
odd Grassmann variables with dimension $n$). Moreover, $|f|$ denotes the grading of $f$.}
\begin{eqnarray}\label{2.5}
H_{_{\mu\nu\rho}}=(-1)^{^{\mu}}\;\frac{\overrightarrow{\partial} B_{_{\nu\rho}}}{\partial
x^{^{\mu}}} +(-1)^{^{\nu+\mu(\nu+\rho)}}\;\frac{\overrightarrow{\partial} B_{_{\rho\mu}}}{\partial
x^{^{\nu}}}+(-1)^{^{\rho(1+\mu+\nu)}}\;\frac{\overrightarrow{\partial}B_{_{\mu\nu}}}{\partial
x^{^{\rho}}}.
\end{eqnarray}

Let us now introduce the action of the YB deformed WZW model on a Lie supergroup $G$ based on solutions of the graded (m)CYBE.
The general procedure that we shall apply is a straightforward generalization of
the deformation prescription invented by Delduc {\it et al.} \cite{Delduc4}. This is basically a two-parameter deformation.

\subsubsection*{{\it \underline{YB deformed WZW action on a Lie supergroup $G$}}}

The action of the YB deformed WZW model on a Lie supergroup $G$ takes the form \cite{Epr2}
\begin{eqnarray}\label{2.6}			
S^{^{YB }}_{_{WZW}}(g)=\frac{1}{2}\int_{_\Sigma}d\sigma^+ d\sigma^-  (-1)^a \mathbb{J}_{+}^{a} \Omega_{_{ab}} L_{-}^{b}  +\frac{\kappa}{12}\
\int_{_{B_{3}}}d^{3}\sigma (-1)^{a+bc}  \varepsilon^{\alpha\beta\gamma}   L_{\alpha}^{a} \Omega_{ad}~ f^d_{~bc} ~ L_{\beta}^{b} L_{\gamma}^{c},~~
\end{eqnarray}
where in the first term the deformed current $\mathbb{J}_{+}$ is defined as follows:
\begin{eqnarray}\label{2.7}
\mathbb{J}_{+} = (1+\omega \eta^{2})\frac{1 + \tilde{A} \mathbb{R}}{1-\eta^{2}\mathbb{R}^{2}} L_{+}.
\end{eqnarray}
Note that this current is a Lie superalgebra-valued one, that is, $\mathbb{J}_{+} =(-1)^a  \mathbb{J}_{+}^{a} T_{_a} $.
Defining $\mathbb{R}(T_{_a})= (-1)^{b}~ \mathbb{R}_a^{~b} T_{_b}$ and $({\mathbb{R}^2})_a^{~b}= (-1)^{c}~ \mathbb{R}_a^{~c} \mathbb{R}_c^{~b}$
it is convenient to write down equation \eqref{2.7} in the following form
\begin{eqnarray}\label{2.8}
\mathbb{J}^a_{+} -(-1)^{b+c}~\eta^{2} J^b_{+} ~\mathbb{R}_b^{~c}~\mathbb{R}_c^{~a}= (1+\omega \eta^{2})\big[L^{a}_{_+} + (-1)^{b}~ \tilde{A} L^{b}_{_+} ~\mathbb{R}_b^{~a}\big].
\end{eqnarray}
According to \eqref{2.7}, the classical action \eqref{2.6} includes three constant parameters $\eta$, $\tilde{A}$ and ${\kappa}$.
This is basically a two-parameter deformation so that the deformation is measured by $\eta$ and $\tilde{A}$.
The last parameter $\kappa$ is regarded as the level.
By setting $\eta= \tilde A=0$ and $\kappa=1$ one concludes that the action \eqref{2.6} becomes the undeformed WZW model.
A key ingredient contained in the current $\mathbb{J}_{+}$ is a constant linear operator $\mathbb{R}: {\G} \rightarrow {\G}$. In the context of
YB deformations, it is supposed that $\mathbb{R}$ should be super skew-symmetric, that is, $<\mathbb{R}(X) , Y>+ <X  , \mathbb{R}(Y)> =0$ for any $X, Y \in \G$, and satisfies the
graded (m)CYBE in operator form
\begin{eqnarray}\label{2.9}
[\mathbb{R}(X) , \mathbb{R}(Y)]-\mathbb{R}\big([\mathbb{R}(X) , Y]+[X , \mathbb{R}(Y)]\big)=\omega [X , Y],
\end{eqnarray}
where $\omega =-1, 1, 0$, corresponding to inhomogeneous non-split, inhomogeneous split, and homogeneous deformations,
respectively. In fact, we say that equation \eqref{2.9} can be generalized to the graded mCYBE if one sets $\omega \neq 0$,
while the case with $\omega =0$ is the homogeneous graded CYBE.
Expanding $X$ as $X= (-1)^{a}~ x^{a} T_{_a}$ and then using the expression $\mathbb{R}(T_{_a})= (-1)^{b}~ \mathbb{R}_a^{~b} T_{_b}$,
the graded (m)CYBE \eqref{2.9} can be rewritten into the following form \cite{Epr2}:
\begin{eqnarray}\label{2.10}
(-1)^k~\mathbb{R}_a^{~c}~f^k_{~cd} \mathbb{R}_b^{~d}-(-1)^b~\mathbb{R}_a^{~c}~f^d_{~cb} \mathbb{R}_d^{~k}-(-1)^a~\mathbb{R}_b^{~c}~f^d_{~ac} \mathbb{R}_d^{~k} = \omega  ~ f^k_{~ab}.
\end{eqnarray}
It would also be useful to obtain the matrix form of the above equation by using the matrix representation of the structure constants,
$f^c_{~ab} =- ({\cal Y}^c)_{ ab}$, giving
\begin{eqnarray}\label{2.11}
(-1)^d~\mathbb{R}~{\cal Y}^k \mathbb{R}^{^{st}}-(-1)^c~\mathbb{R} ({\cal Y}^d \mathbb{R}_d^{~k})- ({\cal Y}^d \mathbb{R}_d^{~k}) \mathbb{R}^{^{st}} =(-1)^k~ \omega {\cal Y}^k,
\end{eqnarray}
where index $d$ in the first term of the left hand side denotes the column of matrix ${\cal Y}^k$,
while index $c$ in the second term corresponds to the row of matrix ${\cal Y}^d$.
Here the superscript ``st'' in $\mathbb{R}^{^{st}}$ stands for the supertranspose \cite{D}.

In addition, the $\mathbb{R}$-operator is connected to a super skew-symmetric CrM $r \in \G \otimes \G$ in the tensorial notation
through the following formula\footnote{Here the inner product is evaluated on the second site of the CrM.}:
\begin{eqnarray}\label{2.12}
\mathbb{R}(X)=<r ~, ~1 \otimes X>,
\end{eqnarray}
with
\begin{eqnarray}\label{2.13}
r=\frac{1}{2}  r^{ab} \big(T_{_a} \otimes T_{_b} -(-1)^{ab}~ T_{_b} \otimes T_{_a}\big)=\frac{1}{2} r^{ab} ~T_{_a} \wedge T_{_b}.
\end{eqnarray}
Consider that Lie superalgebra $\G=\G_{_B} \oplus  \G_{_F}$ of superdimension $(m|n)$ is spanned by the bosonic basis
$\{t_{_i}\}_{i=1}^{m}$ and fermionic ones $\{S_{_\alpha}\}_{\alpha=m+1}^{m+n}$.
Then, CrM can be written as $r= {r}_{_B}^{ij}  ~ t_i \otimes t_j + {r}_{_F}^{\alpha \beta} ~ S_\alpha  \otimes S_\beta$.
Accordingly, the CrM is considered to be even so that it has the matrix representation $r^{ab}$=diag$(r_{_B}, r_{_F})$.
In other words, fermions with bosons cannot be mixed (grading is preserved), namely, $r^{ab} =0$ if $|a| \neq |b|$.
So, in $r^{ab}$ we have always $|a|+|b|=0$.
The other elements such as $\Omega_{ab}$ and $R_a^{~b}$ are also considered similar to $r^{ab}$, that is,
for them we have $|a|+|b|=0$. Using these one can find the relation between matrix representations of $r$ and $\mathbb{R}$ from equation
\eqref{2.12}, giving us \cite{Epr2}
\begin{eqnarray}\label{2.14}
\mathbb{R}_a^{~b}=-(-1)^{ac} ~\Omega_{ac}  ~  r^{cb}.
\end{eqnarray}
Substituting \eqref{2.14} into equation \eqref{2.10}, we arrive at the standard form of the graded (m)CYBE in terms of the CrM \cite{{N.A},{er.crm}}.
\section{YB deformations of the OSP$(1|2)$ WZW model}
\label{Sec.III}

In this section we first introduce the simple Lie superalgebra $osp(1|2)$, and use \eqref{2.3} to build the WZW model based on the OSP$(1|2)$.
Then, we solve the graded (m)CYBE \eqref{2.11} to obtain the
$\mathbb{R}$-operators and inequivalent real CrMs for the $osp(1|2)$.
Using the resulting $\mathbb{R}$-operators and utilizing the action \eqref{2.6} we obtain all YB deformed backgrounds of the OSP$(1|2)$ WZW model.

\subsection{ The OSP$(1|2)$ WZW model}

The simple Lie superalgebra $osp(1|2)$ has superdimension $(3|2)$ possessing three bosonic generators $H, X_{\pm}$ along with two fermionic ones.
The latter are denotes by $Q_{\pm}$. These five generators obey
the following set of non-trivial (anti-)commutation relations \cite{J.z}
\begin{eqnarray}\label{3.1}
&&[H , X_+]=X_+,~~~~~~~~[H , X_-]=-X_-,~~~~~~~[X_+ , X_-]= 2H,~~~~~~[H , Q_{+}]= \frac{1}{2} Q_{+},\nonumber\\
&&{[H , Q_{-}]}= -\frac{1}{2} Q_{-},~~~~[X_+ , Q_{-}]=-Q_{+},~~~~[X_- , Q_{+}]=-Q_{-},~~~~\{Q_{+} , Q_{-}\}= \frac{1}{2} H,\nonumber\\
&&{\{Q_{+} , Q_{+}\}}= \frac{1}{2} X_+,~~~~\{Q_{-} , Q_{-}\}=- \frac{1}{2} X_-.
\end{eqnarray}
Note that $H, X_{\pm}$ generate a $sl(2 , \mathbb{R})$ subalgebra within $osp(1|2)$.
The $osp(1|2)$ possesses a non-degenerate ad-invariant metric $\Omega_{ab}$
which is defined using the inner product in the matrix representation,
\begin{eqnarray}\label{3.2}
\Omega_{ab}=\left( \begin{tabular}{ccccc}
                $\frac{1}{2}$ & 0  &  0  & 0  &0\\
                0 & 0  &  1  & 0 &0\\
                0 & 1  &  0  & 0 & 0 \\
                0 & 0  & 0  &  0 & $\frac{1}{2}$ \\
                0 & 0  &  0  & -$\frac{1}{2}$ &0 \\
                 \end{tabular} \right).
\end{eqnarray}
In order to spell out the action of the WZW model we need to adopt specific coordinates on supergroup
OSP(1|2). Here we shall use the parametrization $g = \alpha g_{_B} \beta$ with $g_{_B} \in $SL$(2 , \mathbb{R})$,
\begin{eqnarray}\label{3.3}
g = \alpha g_{_B} \beta =  e^{\psi Q_{+}} ~ e^{y X_+} e^{\varphi H} e^{u X_-} ~e^{\chi Q_{-}},
\end{eqnarray}
where  $(\varphi , y, u)$ and $(\psi , \chi)$ are the bosonic and fermionic fields, respectively.
Moreover, we need to find the left-invariant Maurer-Cartan super one-forms on the OSP(1|2). To this purpose we use the parametrization \eqref{3.3}.
Then, the $L^{a}_{\pm}$'s take the following form
\begin{eqnarray}
L^{^H}_{\pm} &=& \partial_{\pm} \varphi +2u e^{-\varphi} \partial_{\pm}y + \frac{1}{2} u e^{-\varphi} \psi \partial_{\pm}\psi
+ \frac{1}{2}  e^{-\frac{\varphi}{2}} \chi \partial_{\pm}\psi,\nonumber\\
L^{^{X_+}}_{\pm} &=& e^{-\varphi} \partial_{\pm}y + \frac{1}{4}  e^{-\varphi} \psi \partial_{\pm}\psi, \nonumber\\
L^{^{X_-}}_{\pm} &=&\partial_{\pm} u-u \partial_{\pm} \varphi -u^2 e^{-\varphi} \partial_{\pm}y - \frac{1}{4}  \chi \partial_{\pm}\chi
- \frac{1}{4} u^2 e^{-\varphi} \psi \partial_{\pm}\psi -\frac{1}{2} u e^{-\frac{\varphi}{2}} \chi \partial_{\pm}\psi,\nonumber\\
L^{^{Q_+}}_{\pm} &=&  e^{-\varphi} \chi \partial_{\pm}y - e^{-\frac{\varphi}{2}} \partial_{\pm}\psi - \frac{1}{4}  e^{-\varphi} \psi \chi \partial_{\pm}\psi,\nonumber\\
L^{^{Q_-}}_{\pm} &=& u e^{-\varphi} \chi \partial_{\pm}y + \frac{1}{2} \chi \partial_{\pm} \varphi - \partial_{\pm}\chi
-u e^{-\frac{\varphi}{2}} \partial_{\pm}\psi - \frac{1}{4}  u e^{-\varphi} \psi \chi \partial_{\pm}\psi.\label{3.4}
\end{eqnarray}
Finally by using \eqref{2.3}, the OSP(1|2) WZW action looks like\footnote{In order to calculate the OSP(1|2) WZW action one might use the Polyakov-Wiegmann identity
\begin{align*}
S_{_{_{\hspace{0mm}WZW}}}(g) &=S_{_{_{\hspace{0mm}WZW}}}(g_{_B}) + \int d \sigma^+ d \sigma^-  \big<\alpha^{-1} \partial_{_{-}} \alpha ~
,~ \partial_{_{+}} g_{_B} {g_{_B}}^{-1}\big> \nonumber\\
&+ \int d \sigma^+ d \sigma^-  \big<{g_{_B}}^{-1} \partial_{_{-}} g_{_B} ~
,~ \partial_{_{+}} \beta  {\beta}^{-1}\big>
+\int d \sigma^+ d \sigma^-  \big<\alpha^{-1} \partial_{_{-}} \alpha ~ , ~ g_{_B} (\partial_{_{+}} \beta  {\beta}^{-1}) {g_{_B}}^{-1}\big>.
\end{align*}
With the help of a supergroup element \eqref{3.3},
we did this and the result was exactly the same as what we found in formula \eqref{3.5}.
A derivation of the action \eqref{3.5} can be found in \cite{volker}.
There, in addition to calculating the 2- and 3-point functions of the model, it was shown how local correlation functions in the
OSP(1|2) WZW model can be obtained from those of $N = 1$ supersymmetric Liouville
field theory.}
\begin{align}\label{3.5}
S_{_{_{\hspace{0mm}WZW}}}(g) &= \frac{1}{2} \int d \sigma^+ d \sigma^-\Big[\frac{1}{2} \partial_{_{+}} \varphi \partial_{_{-}} \varphi+ 2  e^{-\varphi}
\partial_{_{+}} u \partial_{_{-}} y+ \frac{1}{2}  e^{-\varphi} \partial_{_{+}} u \psi \partial_{_{-}} \psi\nonumber\\
&~~~~~~~~~~~~~~~~~~~~~~~~~- \frac{1}{2}  e^{-\varphi} \partial_{_{+}} \chi ~\chi  \partial_{_{-}} y+e^{-\frac{\varphi}{2}} \partial_{_{+}} \chi  \partial_{_{-}} \psi +\frac{1}{8}
e^{-\varphi} \partial_{_{+}} \chi ~\psi \chi~ \partial_{_{-}} \psi\Big].
\end{align}
As can be seen from the action, the theory is interacting with terms up to forth order in the fermionic fields.
On the other hand, by regarding this action as a $\sigma$-model action of the form
\eqref{2.4}, we can read off the line element and $B$-field as follows:
\begin{align}
ds^2 = &\frac{1}{2} d\varphi^2+ 2 e^{-\varphi} dydu + \frac{1}{2} e^{-\varphi} \chi ~ dy d\chi + \frac{1}{2} e^{-\varphi}  \psi du d\psi
-(e^{-\frac{\varphi}{2}} + \frac{1}{8} e^{-\varphi} \psi \chi)~ d\psi d\chi,~~~\label{3.6}\\
B = &-e^{-\varphi} ~dy \wedge du - \frac{1}{4} e^{-\varphi} {\chi} ~dy \wedge d\chi +\frac{1}{4} e^{-\varphi} {\psi} ~du \wedge d\psi\nonumber\\
&+\frac{1}{2}  \big(e^{-\frac{\varphi}{2}} + \frac{1}{8} e^{-\varphi} \psi \chi\big) d \psi \wedge d \chi.\label{3.7}
\end{align}
As a WZW model, this model should be conformally invariant. To check this,
one must look at the one-loop beta function equations on supermanifolds \cite{ER5}.
We find that the scalar curvature of the metric is ${\cal R} =-3/4$. Then,
one verifies the one-loop beta function equations with a constant dilaton field.

\subsection{Inequivalent solutions of the graded (m)CYBE}

As mentioned in the Introduction section, in Ref. \cite{J.z}, to calculate Poisson brackets on the OSP$(1|2)$,
three independent complex CrMs were obtained so that they were all coboundaries. In what follows we classify real CrMs for the $osp(1|2)$ as solutions of
the graded (m)CYBE and show that they are split into five inequivalent families.
Before proceeding to solve the graded (m)CYBE \eqref{2.11}, let us assume that
the most general super skew-symmetric CrM $r \in {\cal G}_{_{(3|2)}} \otimes {\cal G}_{_{(3|2)}}$ has the following form:
\begin{eqnarray}\label{3.8}
r=  m_1 H \wedge X_{+} +m_2 H \wedge X_{-} + m_3 X_{+} \wedge X_{-}
+\frac{1}{2} m_4 Q_{+} \wedge Q_{+} +\frac{1}{2} m_5 Q_{-} \wedge Q_{-}+m_6 Q_{+} \wedge Q_{-},
\end{eqnarray}
where $m_i$'s are some real parameters.
Inserting \eqref{3.2} and \eqref{3.8} into \eqref{2.14} one can obtain the general form of the corresponding $\mathbb{R}$-operator. Thus,
by substituting the resulting $\mathbb{R}$-operator and also the matrix representation of the structure constants of \eqref{3.1} into the graded (m)CYBE \eqref{2.11},
one concludes that the general solution is split into two families.
The first family of solutions is given by
\begin{eqnarray}\label{3.9}
{\mathbb{R}_{_I}}_{_a}^{~b}=\left( \begin{tabular}{ccccc}
                0 & $-\frac{1}{2} m_1$  &  $-\frac{1}{2} m_2$  & 0 & 0\\
                $m_2$     & $\pm \sqrt{m_1 m_2}$  &  0        & 0 & 0\\
                $m_1$     & 0  & $\mp \sqrt{m_1 m_2}$  & 0 & 0\\
                0     & 0  & 0 & 0 & 0\\
                0     & 0  & 0 & 0 & 0\\
                 \end{tabular} \right).
\end{eqnarray}
Here the condition \eqref{2.11} has led to the following constraints:
\begin{eqnarray}\label{3.10}
m_3^2=m_1 m_2,~~~~~m_4=0, ~~~~~ m_5=0, ~~~~~ m_6=0,~~~~~ \omega=0.
\end{eqnarray}
The other solution is
\begin{eqnarray}\label{3.11}
{\mathbb{R}_{_{II}}}_{_a}^{~b}=\left( \begin{tabular}{ccccc}
                0 & $-\frac{1}{2} m_1$  &  $-\frac{1}{2} m_2$  & 0 & 0\\
                $m_2$     & $\pm \sqrt{m_1 m_2 -\omega}$  &  0        & 0 & 0\\
                $m_1$     & 0  & $\mp \sqrt{m_1 m_2-\omega}$  & 0 & 0\\
                0     & 0  & 0 &  $\mp \sqrt{m_1 m_2-\omega}$ &  $- m_2$\\
                0     & 0  & 0 &  $m_1$ &  $\pm \sqrt{m_1 m_2-\omega}$\\
                 \end{tabular} \right),
\end{eqnarray}
for which we have the following constraints:
\begin{eqnarray}\label{3.12}
m_3^2=m_1 m_2 - \omega,~~~~m_4= -2 m_1,~~~~m_5=-2 m_2, ~~~~ m_6=-2 m_3.
\end{eqnarray}
Utilizing equations \eqref{2.14} and \eqref{3.2} one can obtain the multiparametric form of CrMs corresponding to the $\mathbb{R}$-operators
${\mathbb{R}_{_{I}}}_{_a}^{~b}$ and ${\mathbb{R}_{_{II}}}_{_a}^{~b}$, giving us
\begin{eqnarray}
{r_{_{I}}}&=& m_1 H \wedge X_+ + m_2 H \wedge X_-  \pm \sqrt{m_1 m_2} ~X_+ \wedge X_-,\label{3.13}\\
{r_{_{II}}}&=& m_1 (H \wedge X_+ - Q_+ \wedge Q_+) + m_2 ~(H \wedge X_- - Q_- \wedge Q_-)\nonumber\\
&&~~~~~~~~~~~~~~~~~~~~~~~~~~~~~~~~\pm  \sqrt{m_1 m_2-\omega}
~(X_+ \wedge X_- - 2  Q_+ \wedge Q_-).\label{3.14}
\end{eqnarray}
One can see that our CrMs are in agreement with those of \cite{J.z} $({r_{_{I}}} \leftrightarrow r_b, {r_{_{II}}} \leftrightarrow r_a)$.
The only difference is in the parameters used in the CrMs, which are considered to be complex in \cite{J.z}, while they are real for us.
As we will show, the realness of the parameters increases the number of inequivalent CrMs.
In this way, one of our main tasks is to determine the exact value of the parameters $m_1$ and $m_2$ in the above CrMs.
Then we can say that we have classified the inequivalent real CrMs for the $osp(1|2)$.
Let us get back to our task.
According to the proposition proved in \cite{Epr1} (see, also, \cite{Epr2}), two CrMs $r$ and $r'$ of a Lie superalgebra $\G$ are equivalent if one can be obtained from the other by means of
a change of basis which is an automorphism $A$ of $\G$, such that
\begin{eqnarray}\label{3.15}
r^{ab} = (-1)^d~ (A^{^{st}})^a_{~c}~ {r^\prime}^{cd}~{A_d}^b.
\end{eqnarray}
Given the above formula, one must obtain the automorphism supergroup of $\G$
which preserves the parity of the generators, as well as the structure constants $f^c_{~ab}$.
Therefore it is crucial for our further considerations to identify the
supergroup of automorphism of the $osp(1|2)$.
We define the action of the automorphism $A$ on $\G$ by the transformation $T'_a = (-1)^b~ {A}_a^{~b}~ T_b$.
The automorphism transformation of  $osp(1|2)$ which preserves the (anti-)commutation rules \eqref{3.1} is
given by
\begin{align}\label{3.16}
H' =&(1+2bc) H +\frac{c(1+bc)}{a} X_+ -ab X_-,\nonumber\\
X'_{+} =&  \frac{2b(1+bc)}{a} H +\frac{(1+bc)^2}{a^2} X_+  -b^2 X_-,\nonumber\\
X'_{-} =&  -{2ac} H -{c^2} X_+  +a^2 X_-,\nonumber\\
Q'_{+} =&- \frac{(1+bc)}{a} Q_{+} -b Q_{-},\nonumber\\
Q'_{-} =&  -c Q_{+} -a Q_{-},
\end{align}
where $a, b, c$ are some arbitrary real numbers.
The bases $\{H', X'_+,  X'_-, Q'_{+},  Q'_{-}\}$ obey the same (anti-)commutation relations as $\{H, X_+,  X_-, Q_{+},  Q_{-}\}$.
When taken into account, the above transformation leads to a conclusion that the parameters $m_{1}$ and $m_{2}$
in \eqref{3.13} and \eqref{3.14} can be scaled out to take the value of $0$ or $1$.
Now, by using the transformation \eqref{3.16} and by employing formula \eqref{3.15}
we arrive at five families of inequivalent CrMs for the $osp(1|2)$ whose representatives
can be described by means of the following Theorem.
\begin{theorem}\label{thm1.A}
Any real CrM of the $osp(1|2)$ Lie superalgebra as a solution of the graded (m)CYBE belongs
just to one of the following five inequivalent classes
\begin{eqnarray*}
{r_{_{i, \epsilon}}}   &=& \epsilon H \wedge X_{_-},~~~~~~~~~~~~~~~~~~~\epsilon =\pm1,\\
{r_{_{ii, \epsilon}}}  &=&\epsilon \big(H \wedge X_{_-} - Q_{-} \wedge  Q_{-}\big), ~~\epsilon =\pm1,\\
{r_{_{\omega}}} &=& H \wedge X_{_+} - Q_{+} \wedge  Q_{+} + \omega \big(H \wedge X_{_-} - Q_{-} \wedge  Q_{-}\big),~~\omega \neq 0.
\end{eqnarray*}
\end{theorem}
\vspace{-5mm}
$\\$
Note that ${r_{_{i, \epsilon}}}$ and ${r_{_{ii, \epsilon}}}$ can also be written in equivalent forms as
$ \epsilon H \wedge X_{_+}$ and $\epsilon \big(H \wedge X_{_+} - Q_{+} \wedge  Q_{+}\big)$, respectively\footnote{The ${r_{_{i, \epsilon=-1}}}$ and
${r_{_{ii, \epsilon=-1}}}$ are inequivalent solutions that are not mentioned in \cite{J.z}.}.
As it stands they satisfy only the graded (not modified) CYBE, whereas the ${r_{_{\omega}}}$ does not.

\subsection{YB deformed backgrounds of the OSP$(1|2)$ WZW model}

Here we present our main results, the YB deformed backgrounds of the OSP$(1|2)$ WZW model.
We proceed to deform the action of the OSP$(1|2)$ WZW model by following the strategy developed in \cite{Epr2}.
In what follows, we use formulas \eqref{2.14} and \eqref{3.2}
to obtain all $\mathbb{R}$-operators corresponding to the inequivalent CrMs of Theorem 1.
Finally, by using the resulting $\mathbb{R}$-operators and also by utilizing relations
\eqref{2.8} and \eqref{3.4} together with the action \eqref{2.6} one can obtain all YB deformed backgrounds of the OSP$(1|2)$ WZW model
including metric and $B$-field.
Notice that the symbol of each background, e.g. OSP$(1|2)_{{i, \epsilon}}^{(\eta,\tilde{A},\kappa)}$,
indicates the deformed background derived by ${r_{_{i, \epsilon}}}$; roman numbers $i$, $ii$ etc.
distinguish between several possible deformed backgrounds of the OSP$(1|2)$ WZW model,
and the $(\kappa, \eta, \tilde{A})$ indicate the deformation parameters of each background.

\subsubsection{Deformation by the ${r_{_{i, \epsilon}}}$: YB deformed backgrounds OSP$(1|2)_{{i, \epsilon=\pm 1}}^{(\eta,\tilde{A},\kappa)}$}
Let us now present the YB deformed backgrounds whose initial inputs are the ${r_{_{i, \epsilon}}}$.
The corresponding $\mathbb{R}$-operator can be obtained by using formulas \eqref{2.14} and \eqref{3.2}, giving us
\begin{eqnarray}\label{3.17}
{\mathbb{R}}_{_a}^{~b}=\left( \begin{tabular}{ccccc}
                0  & 0  &  -$\frac{\epsilon}{2}$  & 0 & 0\\
                ${\epsilon}$  & 0 &  0  & 0 & 0\\
                0  & 0  & 0  & 0 & 0\\
                0  & 0  & 0 & 0 & 0\\
                0  & 0  & 0 & 0 & 0\\
                 \end{tabular} \right).
\end{eqnarray}
In particular, note that this $\mathbb{R}$-operator is nilpotent; $\mathbb{R}^3 =0$.
Then, employing formula \eqref{2.8} and using the fact that $\omega$ is zero we get the corresponding deformed currents
\begin{eqnarray}\label{3.18}
\mathbb{J}^1_{+} &=& L^{1}_{_+} + \epsilon {\tilde A} L^{2}_{_+},~~~~~~~~~~~~~~~~~~~~~~\mathbb{J}^2_{+} = L^{2}_{_+},\nonumber\\
\mathbb{J}^3_{+} &=& - \frac{\epsilon {\tilde A}}{2}  L^{1}_{_+} - \frac{\eta^2}{2} L^{2}_{_+} +L^{3}_{_+},~~~~~~~~\mathbb{J}^4_{+} = L^{4}_{_+},~~~~~~~~\mathbb{J}^5_{+} = L^{5}_{_+}.
\end{eqnarray}
To obtain the explicit form of $\mathbb{J}_{+}$'s one must use \eqref{3.4}. Then, applying the action \eqref{2.6}, the deformed backgrounds including line element and $B$-field are, respectively, given by
\begin{align}
ds_{_{def}}^2 = &\frac{1}{2} d\varphi^2 -\frac{\eta^2}{2} e^{-2 \varphi} ~ dy^2 + 2 e^{-\varphi} dydu -\frac{\eta^2}{4} e^{-2 \varphi} \psi ~ dy d\psi
 + \frac{1}{2} e^{-\varphi} \chi ~ dy d\chi\nonumber\\
 &~~~~~~~~~~~~~~~~~~~~~~~~~~~~~~~~~ + \frac{1}{2} e^{-\varphi}  \psi du d\psi -(e^{-\frac{\varphi}{2}} + \frac{1}{8} e^{-\varphi} \psi \chi)~ d\psi d\chi,~~~\label{3.19}\\
B_{_{def}}= &-\kappa e^{-\varphi} ~dy \wedge du - \frac{\epsilon {\tilde A}}{2} e^{-\varphi}~ d\varphi \wedge dy - \frac{\epsilon {\tilde A}}{8} e^{-\varphi} \psi ~ d\varphi \wedge d\psi
- \frac{\kappa}{4} e^{-\varphi} {\chi} ~dy \wedge d\chi \nonumber\\
&  +\frac{\kappa}{4} e^{-\varphi} {\psi} ~du \wedge d\psi+\frac{\kappa}{2}  \big(e^{-\frac{\varphi}{2}} + \frac{1}{8} e^{-\varphi} \psi \chi\big) d \psi \wedge d \chi +\frac{\epsilon {\tilde A}}{4} e^{-\frac{3 \varphi}{2}} \chi~ dy \wedge d\psi \nonumber\\
&-\frac{\epsilon {\tilde A}}{16} e^{-\frac{3 \varphi}{2}} \psi \chi~ d\psi \wedge d\psi.\label{3.20}
\end{align}
As can be seen, the effect coming from the deformation is reflected as the coefficient of both metric and $B$-field.
The scalar curvature of the above metric is similar to that of the undeformed WZW model, namely, ${\cal R} =-3/4$.
In order to check the conformal invariance conditions of the deformed background we look at
the one-loop beta function equations \cite{ER5}. These equations are satisfied provided that the deformation parameters $\eta$ and
${\tilde A}$ become zero and ${\kappa} =1$, in which case the deformed background turns into the
undeformed one (equations \eqref{3.6} and \eqref{3.7}).

\subsubsection{Deformation by the ${r_{_{ii, \epsilon}}}$: YB deformed backgrounds OSP$(1|2)_{{ii, \epsilon=\pm 1}}^{(\eta,\tilde{A},\kappa)}$}

For the CrM  ${r_{_{ii, \epsilon}}}$, one can get the corresponding $\mathbb{R}$-operator by using formulas \eqref{2.14} and \eqref{3.2}.
Then, using \eqref{2.8} we find that
\begin{eqnarray}\label{3.21}
\mathbb{J}^1_{+} &=& L^{1}_{_+} + \epsilon {\tilde A} L^{2}_{_+},~~~~~~~~~~~~~~~~~~~~~~\mathbb{J}^2_{+} = L^{2}_{_+},\nonumber\\
\mathbb{J}^3_{+} &=& - \frac{\epsilon {\tilde A}}{2}  L^{1}_{_+} - \frac{\eta^2}{2} L^{2}_{_+} +L^{3}_{_+},~~~~~~~~
\mathbb{J}^4_{+} = L^{4}_{_+},~~~~~~~~\mathbb{J}^5_{+} = \epsilon {\tilde A} L^{4}_{_+} +L^{5}_{_+}.
\end{eqnarray}
Finally, inserting \eqref{3.4} into \eqref{3.21} and then applying \eqref{2.6}, the deformed backgrounds read
\begin{align}
ds_{_{def}}^2 = &\frac{1}{2} d\varphi^2 -\frac{\eta^2}{2} e^{-2 \varphi} ~ dy^2 + 2 e^{-\varphi} dydu -\frac{\eta^2}{4} e^{-2 \varphi} \psi ~ dy d\psi
 + \frac{1}{2} e^{-\varphi} \chi ~ dy d\chi\nonumber\\
 &~~~~~~~~~~~~~~~~~~~~~~~~~~~~~~~~~ + \frac{1}{2} e^{-\varphi}  \psi du d\psi -(e^{-\frac{\varphi}{2}} + \frac{1}{8} e^{-\varphi} \psi \chi)~ d\psi d\chi,~~~\label{3.22}\\
B_{_{def}}= &-\kappa e^{-\varphi} ~dy \wedge du - \frac{\epsilon {\tilde A}}{2} e^{-\varphi}~ d\varphi \wedge dy - \frac{\epsilon {\tilde A}}{8} e^{-\varphi} \psi ~ d\varphi \wedge d\psi
- \frac{\kappa}{4} e^{-\varphi} {\chi} ~dy \wedge d\chi \nonumber\\
& +\frac{\kappa}{4} e^{-\varphi} {\psi} ~du \wedge d\psi+\frac{\kappa}{2}  \big(e^{-\frac{\varphi}{2}} + \frac{1}{8} e^{-\varphi} \psi \chi\big) d \psi \wedge d \chi -\frac{\epsilon {\tilde A}}{4} e^{-\frac{3 \varphi}{2}} \chi~ dy \wedge d\psi \nonumber\\
&+\frac{\epsilon {\tilde A} e^{-\varphi} }{4}\big(1 +\frac{1}{4} e^{-\frac{ \varphi}{2}} \psi \chi\big)~ d\psi \wedge d\psi.\label{3.23}
\end{align}
Note that the deformed background above is similar to the previous one, equations \eqref{3.19} and \eqref{3.20}.
The only difference is in the last two terms of the $B$-field.
In the case of the check of conformal invariance, we also encounter conditions similar to the previous case.

\subsubsection{Deformation by the ${r_{_{\omega}}}$: YB deformed background OSP$(1|2)_{{\omega}}^{(\eta,\tilde{A},\kappa)}$}

The $\mathbb{R}$-operator corresponding to the CrM ${r_{_{\omega}}}$ is
\begin{eqnarray}\label{3.24}
{\mathbb{R}}_{_a}^{~b}=\left( \begin{tabular}{ccccc}
                0  & -$\frac{1}{2}$  &  -$\frac{\omega}{2}$  & 0 & 0\\
                ${\omega}$  & 0 &  0  & 0 & 0\\
                1  & 0  & 0  & 0 & 0\\
                0  & 0  & 0 & 0 & -${\omega}$\\
                0  & 0  & 0 & 1 & 0\\
                 \end{tabular} \right),
\end{eqnarray}
from which we can read off the deformed currents
\begin{align}\label{3.25}
\mathbb{J}^1_{+} =& L^{1}_{_+} + {\tilde A} \big(\omega L^{2}_{_+}+L^{3}_{_+}\big),~~~~~~~~~~~~~~~~~~~~~~~~~~~\mathbb{J}^2_{+} = -\frac{{\tilde A}}{2} L^{1}_{_+} +(1+\frac{\omega {\eta^2}}{2}) L^{2}_{_+} -\frac{{\eta^2}}{2} L^{3}_{_+},\nonumber\\
\mathbb{J}^3_{+} =& -\frac{\omega {\tilde A}}{2} L^{1}_{_+}-  \frac{{\omega^2 \eta^2}}{2} L^{2}_{_+} +(1+\frac{\omega {\eta^2}}{2}) L^{3}_{_+},~~~~~
\mathbb{J}^4_{+} = L^{4}_{_+} - {\tilde A} L^{5}_{_+},~~~~\mathbb{J}^5_{+} = \omega {\tilde A} L^{4}_{_+} +L^{5}_{_+}.
\end{align}
Finally, the background looks like
\begin{align}
ds_{_{def}}^2 = &\frac{1}{2}(1-\eta^2 u^2) d\varphi^2 -\frac{\eta^2}{2} (\omega +u^2)^2 e^{-2 \varphi} ~ dy^2 -\frac{\eta^2}{2} du^2+{\eta^2} u ~d\varphi du -{\eta^2} (\omega +u^2) u e^{-\varphi} ~ d\varphi dy\nonumber\\
&+\big[2+ (\omega +u^2) \eta^2\big] e^{-\varphi} dydu  + \frac{1}{2} \big[\big(1 +\frac{\eta^2}{2}(\omega +u^2)\big) e^{-\varphi}  \psi + {\eta^2} e^{-\frac{\varphi}{2}} u \chi\big] du d\psi
 \nonumber\\
&+ \frac{1}{2}\big[1-\frac{\eta^2}{2} (\omega +u^2)\big]e^{-\varphi} \chi ~ dy d\chi-\frac{\eta^2}{4} (\omega +u^2)\big[(\omega +u^2) e^{-2 \varphi} \psi +2e^{-\frac{3\varphi}{2}} u \chi\big]~ dy d\psi  \nonumber\\
&-\frac{\eta^2}{2} \big[u e^{-\frac{\varphi}{2}} \chi +\frac{1}{2} (\omega +u^2)e^{-\varphi} \psi\big] u~ d\varphi d\psi
  -\frac{\eta^2}{4} u \chi ~d\varphi d \chi +\frac{\eta^2}{4} \chi~ du d\chi  \nonumber\\
&-\Big[(e^{-\frac{\varphi}{2}} + \frac{1}{8} e^{-\varphi} \psi \chi)-\frac{\eta^2}{16}(\omega +u^2)  e^{-\varphi} \psi \chi\Big]~ d\psi d\chi,\label{3.26}\\
B_{_{def}}= & - \frac{{\tilde A}}{2} (\omega +u^2) e^{-\varphi}~ d\varphi \wedge dy
- \frac{{\tilde A}}{2} ~ d\varphi \wedge du -\frac{{\tilde A}}{4} \Big[u e^{-\frac{\varphi}{2}} \chi +\frac{1}{2} e^{-\varphi}(\omega +u^2) \psi \Big] ~ d\varphi \wedge d\psi\nonumber\\
&  -(\kappa+ {\tilde A} u) e^{-\varphi} ~ dy \wedge du
-\frac{{\tilde A}}{4} (\omega +u^2) e^{-\frac{3 \varphi}{2}}  \chi~ dy \wedge d\psi
-\frac{1}{4} ({\kappa} + {\tilde A} u)e^{-\varphi} {\chi} ~dy \wedge d\chi \nonumber\\
&- \frac{{\tilde A}}{8} \chi~  d\varphi \wedge d\chi +\frac{1}{4}\big[\kappa e^{-\varphi} {\psi} +{\tilde A} (u e^{-\varphi} \psi+ e^{-\frac{\varphi}{2}} \chi )\big] ~du \wedge d\psi + \frac{{\tilde A}}{4} ~ d\chi \wedge d \chi \nonumber\\
&+\frac{1}{2}  (\kappa + {\tilde A} u)  \big(e^{-\frac{\varphi}{2}} + \frac{1}{8} e^{-\varphi} \psi \chi\big) d \psi \wedge d \chi
+\frac{{\tilde A}}{4} (\omega +u^2) \big(e^{-\varphi}  +\frac{1}{4} e^{-\frac{ \varphi}{2}} \psi \chi\big)~ d\psi \wedge d\psi.\label{3.27}
\end{align}
In the next section, we discuss the above deformed solutions in the context of graded generalized supergravity equations.
In particular, the undeformed WZW background is also considered.
\section{Non-unimodularity and solutions of graded generalized supergravity equations}
\label{Sec.IV}

In this section, we look at the (non-)Abelian and  (non-)unimodularity conditions on the CrMs of the $osp(1|2)$.
We then investigate whether the deformed backgrounds generated by non-unimodular CrMs can be solutions to graded generalized supergravity equations.

As mentioned in \cite{K.Yoshida}, the target spacetime of YB deformations of $AdS_5 \times S^5$ based on the homogeneous CYBE \cite{{Delduc2},{Delduc3}}
satisfies the equations of motion of type IIB supergravity if
the CrM satisfies the unimodularity condition \cite{Wulff1}. If not, the background is a
solution of so-called generalized type IIB supergravity \cite{Arutyunov3,Wulff2}.
So, non-unimodular YB deformations result in solutions of generalized supergravity.
As we will show, the last sentence does not work for our backgrounds.
As previously mentioned, the CrMs are divided into Abelian and non-Abelian.
The CrM is called Abelian if $[T_a , T_b] =0$, otherwise, it is non-Abelian. Also, it is called unimodular if it satisfies the following condition
\begin{eqnarray}
r^{ab}~ [T_a , T_b] =0.\label{4.1}
\end{eqnarray}
Clearly, all Abelian CrMs are unimodular. Using \eqref{3.1} together with \eqref{4.1} we find that all CrMs of the $osp(1|2)$ (${r_{_{i, \epsilon}}}, {r_{_{ii, \epsilon}}}$ and ${r_{_{\omega}}}$) are non-Abelian and also non-unimodular.
In the following, we want to answer the question whether the deformed backgrounds generated by non-unimodular CrMs can be solutions to the graded generalized supergravity equations or not?
Recently, we have written the generalized supergravity equations introduced by Arutyunov {\it et al.} \cite {Arutyunov3} on a supermanifold $\M$.
We have called them the graded generalized supergravity equations \cite{e.gh.r4}.
In the absence of the Ramond-Ramond fields, the set of graded generalized supergravity equations on a supermanifold $\M$ with the supercoordinates\footnote{See footnote 3.} $x^\mu$ can be written in the following form \cite{e.gh.r4}
\begin{eqnarray}
{\cal R}_{_{\mu \nu}}+\frac{1}{4} H_{_{\mu \rho\sigma}} {H^{\sigma \rho}}_{\nu}+{\overrightarrow{\nabla}}_\mu X_\nu +(-1)^{\mu \nu}~{\overrightarrow{\nabla}}_\nu X_\mu  &=&0,\label{4.2}\\
\frac{1}{2}(-1)^{\lambda}~{\overrightarrow{\nabla}}^\lambda H_{_{\lambda \mu \nu}} - (-1)^{\lambda} X^{\lambda} H_{_{\lambda \mu \nu}}-
{\overrightarrow{\nabla}}_\mu X_\nu +(-1)^{\mu\nu}{\overrightarrow{\nabla}}_\nu X_\mu &=&0,\label{4.3}\\
-4 \Lambda+{\cal R}+\frac{1}{12}  H_{_{\mu \nu \rho}} H^{^{\rho \nu \mu}} +4 {\overrightarrow{\nabla}}_\mu X^\mu -4 X_\mu X^\mu&=&0,\label{4.4}\\
{\cal L}_{_{I}} G_{_{\mu \nu}} = (-1)^{\lambda +\mu} \frac{\overrightarrow{\partial}I^{\lambda}}{{\partial} x^{^{\mu}}} G_{_{\lambda \nu}}+ I^{\lambda}
\frac{\overrightarrow{\partial} G_{_{\mu\nu}}}{{\partial} x^{^{\lambda}}} + (-1)^{\mu \nu +\mu\lambda + \lambda+ \nu} \frac{\overrightarrow{\partial}I^{\lambda}}{{\partial} x^{^{\nu}}} G_{_{\mu\lambda}}
&=&0, \label{4.5}\\
{\cal L}_{_{I}} B_{_{\mu \nu}}  = (-1)^{\lambda +\mu} \frac{\overrightarrow{\partial}I^{\lambda}}{{\partial} x^{^{\mu}}} B_{_{\lambda \nu}}+ I^{\lambda} \frac{\overrightarrow{\partial} B_{_{\mu\nu}}}{{\partial} x^{^{\lambda}}} + (-1)^{\mu \nu +\mu\lambda + \lambda+ \nu} \frac{\overrightarrow{\partial}I^{\lambda}}{{\partial} x^{^{\nu}}} B_{_{\mu\lambda}}
&=&0, \label{4.6}
\\
{\overrightarrow{\nabla}}_\mu Z_\nu -(-1)^{\mu \nu}~{\overrightarrow{\nabla}}_\nu Z_\mu  + (-1)^{\lambda}I^{\lambda} H_{_{\lambda \mu \nu}}&=&0, \label{4.7}\\
(-1)^{\lambda} I^{\lambda} Z_\lambda&=&0,\label{4.8}
\end{eqnarray}
where ${\cal R}_{_{\mu \nu}}$ and ${\cal R}$ are the respective Ricci tensor and scalar curvature that are calculated from the metric ${G}_{_{\mu \nu}}$,
and $\Lambda$ is the cosmological constant. The covariant superderivative
${\overrightarrow{\nabla}}_\mu $ is the conventional Levi-Civita connection associated with ${G}_{_{\mu \nu}}$\footnote{If
$X=X_\mu dx^\mu$ be a covariant supervector field, then one finds covariant superderivative in explicit components form as follows \cite{D}:
$${\overrightarrow{\nabla}}_\mu  X_\nu =(-1)^{\mu} \frac{\overrightarrow{\partial}X_{\nu}}{{\partial} x^{^{\mu}}} - X_{\sigma} \Gamma^{\sigma}_{\mu\nu},$$
where $\Gamma^{\sigma}_{\mu\nu}$ are called the components of the connection ${\overrightarrow{\nabla}}_\mu $.}.
As can be seen, a supervector field $I=I^{\mu} \frac{\overrightarrow{\partial}}{{\partial} x^{^{\mu}}}$  and a super one-form $Z=Z_\mu dx^\mu$  are defined so as to satisfy equations \eqref{4.5}-\eqref{4.8}.
Moreover, super one-form $X=X_{_\mu} dx^\mu$ is related to the $Z$ via $X_\mu= I_\mu + Z_\mu$, and the conventional dilaton $\Phi$ is included in $Z_\mu$  as follows:
\begin{eqnarray}\label{4.9}
Z_\mu=  (-1)^{\mu} \frac{\overrightarrow{\partial} \Phi }{{\partial} x^{^{\mu}}} + (-1)^{\lambda }  I^{\lambda}  B_{_{\lambda \mu}}.
\end{eqnarray}
A remarkable point is that if one sets $I^{\lambda} = 0$, then we find that $X_\mu= (-1)^{\mu} {\overrightarrow{\partial} \Phi}/{{\partial} x^{^{\mu}}}$, and thus, the graded generalized supergravity equations reduce to the standard supergravity equations (corresponding to the one-loop beta function equations in
$\sigma$-model language \cite{callan}).

In order to examine the above equations for the deformed backgrounds, we first focus on equations \eqref{4.5} and \eqref{4.6}.
By combining these two equations, it would conclude that ${\cal L}_{_{I}} {\mathcal{E}}_{_{\mu \nu}}=0 $, where
the background field ${\mathcal{E}}_{_{\mu \nu}}$ can be understood as a sum of the metric $G_{_{\mu \nu}}$ and field $B_{_{\mu \nu}}$.
Indeed, the aforementioned equation is nothing but the Killing equation.
Accordingly, the supervector field $I$ can be a Killing supervector or
a linear combination of the Killing supervectors corresponding
to the field ${\mathcal{E}}_{_{\mu \nu}}$. In fact, to construct an appropriate supervector field one may choose
\begin{eqnarray}\label{4.10}
I=\sum_{i=1}^{m} c_{_i} K^B_{_i} +\sum_{\alpha=1}^{n} a_{_\alpha} K^F_{_\alpha},
\end{eqnarray}
where $m$ and $n$ stand for the number of bosonic and fermionic Killing vectors of the background field. Moreover,
$c_{_i}$'s and  $a_{_\alpha}$'s are  $c$- and $a$-numbers \cite{D}, respectively.
For a Killing supervector field $K_a={K_a}^{\mu} \frac{\overrightarrow{\partial}}{{\partial} x^{^{\mu}}}$, the Killing equation can be written as
\begin{eqnarray}\label{4.11}
{\cal L}_{_{K_a}} {\mathcal{E}}_{_{\mu \nu}} = (-1)^{\mu +\lambda +\mu a} \frac{\overrightarrow{\partial} {K_a}^{\lambda}}{{\partial} x^{^{\mu}}} {\mathcal{E}}_{\lambda \nu}+ {K_a}^{\lambda} \frac{\overrightarrow{\partial} {\mathcal{E}}_{\mu\nu}}{{\partial} x^{^{\lambda}}}
+ (-1)^{\mu \nu +\mu\lambda + \lambda+ \nu a +\nu} \frac{\overrightarrow{\partial} {K_a}^{\lambda}}{{\partial} x^{^{\nu}}} {\mathcal{E}}_{\mu\lambda} = 0.
\end{eqnarray}
The Killing supervector $K_a$ is considered to have two degrees, even (bosonic) and odd (fermionic).
Now we assume that ${K_a}^{ \mu}$ is in the form of a $2 \times 2$ block matrix as
\begin{equation}\label{4.12}
{K_a}^{~\mu}\;=\;\left(
\begin{tabular}{c|c}
                 ${K^{^B}_{\;i}}^{\;\mu_i}$ & ${K^{^B}_{\;i}}^{\;\mu_\alpha}$ \\
\hline
                 ${K^{^F}_{\;\alpha}}^{\;\mu_i}$ & ${K^{^F}_{\;\alpha}}^{\;\mu_\alpha}$ \\
                 \end{tabular} \right),
\end{equation}
where the elements of the submatrices ${K^{^B}_{~i}}^{\;\mu_i}$ and ${K^{^F}_{\; \alpha}}^{\;\mu_\alpha}$ are even ($c$-numbers) such that
$|i| +|{\mu_i}| = |\alpha| +|\mu{_\alpha}|=0$, whereas  the elements ${K^{^B}_{\;i}}^{\;\mu_\alpha}$ and  ${K^{^F}_{\;\alpha}}^{\;\mu_i}$ are odd
($a$-numbers) for which $|i| +|\mu{_\alpha}| = |\alpha| +|{\mu_i}|=1$.
According to the definitions above, the supervector $I$ is considered to be even.

Before proceeding to solve equations \eqref{4.2}-\eqref{4.8} for the deformed backgrounds,
let us examine the equations for the undeformed OSP$(1|2)$ WZW background.
It seems interesting to show that the OSP$(1|2)$ WZW model satisfies the graded generalized supergravity equations.

\subsection{The OSP$(1|2)$ WZW model and the graded generalized supergravity equations}

In this subsection we show that the OSP$(1|2)$ WZW background can be considered as a solution to the graded generalized supergravity equations.
We shall achieve this with the help of formula \eqref{4.11}.
The background field ${\mathcal{E}}_{_{\mu \nu}}$ corresponding to the model can be obtained from equations \eqref{3.6} and \eqref{3.7}.
Then, using equation \eqref{4.11} we obtain the corresponding Killing supervectors.
The background admits four bosonic Killing vectors
\begin{align}\label{4.13}
K^B_{_1} = 2\overrightarrow{\frac{\partial}{\partial \varphi}} + 2 y \overrightarrow{\frac{\partial}{\partial y}} + \psi \overrightarrow{\frac{\partial}{\partial \psi}},~~~~K^B_{_2} = 2\overrightarrow{\frac{\partial}{\partial \varphi}} + 2u \overrightarrow{\frac{\partial}{\partial u}} + \chi \overrightarrow{\frac{\partial}{\partial \chi}},~~~~~~
K^B_{_3} = \overrightarrow{\frac{\partial}{\partial y}},~~~~~~K^B_{_4} = \overrightarrow{\frac{\partial}{\partial u}}.
\end{align}
In addition, there exist two fermionic Killing vectors $K^F_{_1}$ and $K^F_{_2}$ which generate the isometric Lie superalgebra of the background
\begin{align}\label{4.14}
K^F_{_1} =- \frac{\psi }{4} \overrightarrow{\frac{\partial}{\partial y}} +  \overrightarrow{\frac{\partial}{\partial \psi}},~~~~~~~~~~~~~~
K^F_{_2} = - \frac{\chi}{4} \overrightarrow{\frac{\partial}{\partial u}} + \overrightarrow{\frac{\partial}{\partial \chi}}.
\end{align}
By substituting \eqref{4.13} and \eqref{4.14} into formula \eqref{4.10}, we obtain that
\begin{align}\label{4.15}
I = 2(c_{_1}+c_{_2}) \frac{\overrightarrow{\partial} }{{\partial} \varphi}+(c_{_3} +2y c_{_1} - \frac{1}{4} a_{_1} \psi) \frac{\overrightarrow{\partial} }{{\partial} y}&+(c_{_4} +2u c_{_2} - \frac{1}{4} a_{_2} \chi) \frac{\overrightarrow{\partial} }{{\partial} u} \nonumber\\
&+(c_{_1} \psi + a_{_1})\overrightarrow{\frac{\partial}{\partial \psi}} +(c_{_2} \chi + a_{_2})
\overrightarrow{\frac{\partial}{\partial \chi}}.
\end{align}
Here $(c_{_1}, \cdots, c_{_4})$ and  $(a_{_1}, a_{_2})$ are  $c$- and $a$-numbers, respectively.
Indeed, supervector field \eqref{4.15} satisfies equations \eqref{4.5} and \eqref{4.6}.
The remaining equations are satisfied provided that $c_{_1}= c_{_3}=0$ and $a_{_1}=0$, as well as $\Lambda =-1/8-2 {c_{_2}}^2$. Thus, the graded generalized supergravity equations
are satisfied with the background of the OSP$(1|2)$ WZW model together with
\begin{align}\label{4.16}
I &= 2 c_{_2} \frac{\overrightarrow{\partial} }{{\partial} \varphi}+(c_{_4} +2u c_{_2} - \frac{1}{4} a_{_2} \chi) \frac{\overrightarrow{\partial} }{{\partial} u}+(c_{_2} \chi + a_{_2}) \frac{\overrightarrow{\partial} }{{\partial} \chi},\nonumber\\
X&= c_{_2} d \varphi +e^{-\varphi}\big(c_{_4}+2u c_{_2} -\frac{1}{2} a_{_2} \chi\big) dy\nonumber\\
&~~~~~~~~~~~~~~~~~~~~+\frac{1}{2}\Big[e^{\frac{-\varphi}{2}} (c_{_2} \chi + a_{_2})
+ e^{{-\varphi} } (u c_{_2} \psi +\frac{1}{2}c_{_4} \psi +\frac{1}{4} a_{_2} \psi \chi)\Big] d \psi,\nonumber\\
Z&=0.
\end{align}

\subsection{The YB deformed backgrounds OSP$(1|2)_{{i, \epsilon=\pm 1}}^{(\eta,\tilde{A},\kappa)}$ and the graded generalized supergravity equations}
The backgrounds OSP$(1|2)_{{i, \epsilon}}^{(\eta,\tilde{A},\kappa)}$ have fewer symmetries than the undeformed background.
They admit the following Killing supervectors
\begin{align}\label{4.17}
K^B_{_1} =&\overrightarrow{\frac{\partial}{\partial u}} ,~~~~~~~~~~~
K^B_{_2} = 2\overrightarrow{\frac{\partial}{\partial \varphi}} + 2 y \overrightarrow{\frac{\partial}{\partial y}} + \psi \overrightarrow{\frac{\partial}{\partial \psi}},~~~~~~~~~
K^B_{_3} = \overrightarrow{\frac{\partial}{\partial y}},\nonumber\\
K^F_{_1} =&-\frac{1}{4} \psi \overrightarrow{\frac{\partial}{\partial y}} +\overrightarrow{\frac{\partial}{\partial \psi}}.
\end{align}
The corresponding supervector field is then obtained to be
\begin{align}\label{4.18}
I = 2 c_{_2} \frac{\overrightarrow{\partial} }{{\partial} \varphi}+(c_{_3} +2y c_{_2} - \frac{1}{4} a_{_1} \psi) \frac{\overrightarrow{\partial} }{{\partial} y}+ c_{_1} \frac{\overrightarrow{\partial} }{{\partial} u}
&+(c_{_2} \psi + a_{_1})\overrightarrow{\frac{\partial}{\partial \psi}}.
\end{align}
It is easy to check that this supervector satisfies equations \eqref{4.5} and \eqref{4.6}.
However, we need to find suitable super one-forms $Z$ and $X$ so that equations \eqref{4.2}-\eqref{4.4} and \eqref{4.7} together with \eqref{4.8}
are also satisfied.
After some algebraic calculations, we concluded that there is no suitable super one-form $Z$ corresponding to supervector field
\eqref{4.18} that satisfies equations \eqref{4.2}-\eqref{4.8}.
Note that for the remaining backgrounds OSP$(1|2)_{{ii, \epsilon=\pm 1}}^{(\eta,\tilde{A},\kappa)}$ and OSP$(1|2)_{{\omega}}^{(\eta,\tilde{A},\kappa)}$
we performed similar calculations as above.
Finally, we see that the deformed OSP$(1|2)$ $\sigma$-models do not satisfy the graded generalized supergravity equations.
As mentioned in the Introduction section, this
is because the deformed OSP$(1|2)$ $\sigma$-models are not of the Green-Schwarz type,
and therefore fall outside the scope of the Borsato-Wulff theorem \cite{Wulff1}.
However, our results are in full agreement with the prior observation made by Bielli {\it et al.} \cite{Bielli1}
in the context of non-Abelian T-duality for the same OSP$(1|2)$ $\sigma$-model.

\section{\label{Sec.V} Conclusions}

In this paper, we have studied the integrable deformations of the OSP$(1|2)$ WZW model.
Our strategy here was to follow a prescription invented by Delduc {\it et al.} \cite{Delduc4},
which was generalized to a Lie supergroup case in \cite{Epr2}.
This is basically a two-parameter deformation.
Here we extend their prescription to the OSP$(1|2)$ WZW model.
By considering the most general super skew-symmetric CrM, we have obtained five inequivalent families of CrMs for the $osp(1|2)$ Lie superalgebra as real solutions of the graded (m)CYBE.
All the resulting CrMs were non-Abelian and also non-unimodular.
In this regard, we were able to obtain a family of integrable $\sigma$-models called the YB deformations of the OSP$(1|2)$ WZW model.
We have shown that the effect coming from the deformation is reflected as the coefficient of both metric and $B$-field.
The non-unimodular CrMs of the $osp(1|2)$ did not lead us to the graded generalized supergravity equations. 
In fact, our deformed backgrounds could not satisfy the graded generalized equations.
The fact that our deformed OSP$(1|2)$ $\sigma$-models did not satisfy these equations is consistent with expectations.
We should note that this result does not contradict the result of \cite{Wulff1} since the deformed models under consideration do not describe a
Green-Schwarz superstring.
Anyway, the deformed backgrounds that we found are interesting, in particular in the OSP$(1|2)$ case they are rare, much hard technical work was needed to obtain them.

\subsection*{Declaration of competing interest}

The authors declare that they have no known competing financial
interests or personal relationships that could have appeared to influence the work reported in this paper.

\subsection*{Acknowledgements}

The authors would like to thank the anonymous referee for invaluable comments and criticisms.
This work has been supported by the research vice chancellor of Azarbaijan Shahid Madani University under research fund No. 1403/537.


\subsection*{Data availability statement}

No data was used for the research described in the article.
\\
\\
{\bf ORCID iDs}
\\
Ali Eghbali ~  https://orcid.org/0000-0001-6076-2179
\\
Yaghoub Samadi ~ https://orcid.org/0009-0002-4709-3374
\\
Adel Rezaei-Aghdam ~ https://orcid.org/0000-0003-4754-7911


\end{document}